\documentclass[11pt]{article}
\usepackage{epsf,amsmath,amssymb,afterpage}
\begin{document}

\begin{flushright}
LPTHE Orsay 98/22 \\
BI-TH-98/08 \\
April 1998
\end{flushright}

\begin{center}
\vspace{24pt}

{\Large \bf 4d Simplicial Quantum Gravity: Matter Fields \\
\vspace{5pt} and the Corresponding Effective Action}
\vspace{24pt}

{\large \sl S.~Bilke $^a$, Z.~Burda $^b$, A.~Krzywicki $^c$,
 B.~Petersson $^d$, \\
 J.~Tabaczek $^d$ \,{\rm and}\, G.~Thorleifsson $^d$}\\
\vspace{10pt}
$^a$ Inst. Theor. Fysica, Univ. Amsterdam, 1018 XE Amsterdam,
The Netherlands\\
$^b$ Institute of Physics, Jagellonian
University, 30059 Krakow, Poland\\
$^c$ LPTHE, B\^{a}timent 211, Universit\'e Paris-Sud, 91405 Orsay,
France\footnote{Laboratoire associ\'e au C.N.R.S.}\\
$^d$ Fakult\"{a}t f\"{u}r Physik, Universit\"{a}t Bielefeld,
 33501 Bielefeld, Germany\\
\vspace{10pt}

\begin{abstract} 
Four-dimensional simplicial quantum gravity is modified
either by coupling it to $U(1)$ gauge fields or by
introducing a measure weighted by the orders of the triangles.
Strong coupling expansion and Monte Carlo simulations are used.
Although the two modifications of the standard pure-gravity
model are apparently very distinct, they produce strikingly 
similar results, as far as the geometry of random manifolds 
is concerned. In particular, for an appropriate choice of 
couplings, the branched polymer phase is 
replaced by a {\em crinkled} phase, characterized by the susceptibility 
exponent $\gamma < 0$  and the fractal dimension $d_H > 2$. 
The quasi-equivalence between the two models is exploited 
to get further insight into the extended phase diagram of 
the theory. 
\end{abstract}
\end{center}
\vspace{15pt}

\section{Introduction}

\noindent
As this work follows a couple of other papers which we have devoted 
to the study of the phase diagram of 4d simplicial gravity 
\cite{us1,us2}, we shall not expand much on our motivations. 
Instead, in order to save space, we shall enter without further 
ado into the main body of the work. Let us only observe that following 
the surprising discovery \cite{us1,other} that the phase transition 
between the so-called crumpled and branched polymer phases in 4d 
simplicial gravity is discontinuous, several groups have tried to 
figure out what kind of modification could render the theory more 
realistic. One track, explored up to now in 3d only \cite{syr,hotta}, 
consists in modifying the measure in the partition function following 
a recipe proposed some time ago in Ref.~\cite{bm}. Loosely speaking, 
this corresponds to the introduction of a $R^2$ term in the 
action\footnote{In this text the expression ``$R^2$ term''
refers to any combination of terms quadratic in Riemann, 
not only to the square of the scalar curvature. For early simulations
with $R^2$ term in the action see also 
\cite{jan,hin}.}. Another idea consists 
in introducing matter fields \cite{us2}. Indeed, in the continuum 
formalism, one can argue that {\em adding} conformal matter fields 
in 4d has the effect similar to that obtained by {\em reducing}  
their number in 2d \cite{jk} and might therefore bring one from the 
branched polymer to a physical phase (analogous to the Liouville 
phase in 2d).

In Ref.~\cite{us2} the effect of coupling non-compact $U(1)$ 
gauge fields to 4d simplicial quantum gravity was studied. 
It was found that for more than two gauge fields the 
back-reaction of matter on geometry is strong, that the degeneracy 
of random manifolds into branched polymers does not occur and that 
there is some evidence for a new ''smoother`` phase of the geometry,
to be called hereafter the {\em crinkled} phase.
This apparently confirmed the speculation put forward in \cite{jk}. 
The results were obtained from the strong coupling series, 
whose introduction in this context has been quite novel, 
and from Monte Carlo simulations with lattices of relatively modest
size. An eventual extension of the study to larger systems 
was announced.

We have extended our Monte Carlo simulations to systems with up 
to 32K simplexes, for 3 copies of gauge fields. 
We have also calculated more terms of the strong coupling series. 
While our results are still consistent with the scenario
described above, we do observe some disturbing inconsistencies
in the behavior of different geometric observables which, 
taken at face value, seem incompatible with the existence
of a single phase transition at a finite value of the
Newton coupling $\kappa_2$. 

We have noticed, on the other hand, a curious 
quasi-equivalence between the model with gauge fields and 
that with a properly modified measure. We have concentrated on
this issue, achieving a better understanding of the results 
obtained earlier, and we have explored further both the phase
structure of the model and the nature of the hypothetical crinkled 
phase. All these developments will be described in the 
following sections.

\section{Gauge fields on a random manifold}

\noindent
The action is a sum of two parts. The first is the Einstein-Hilbert
action, which for a 4d simplicial manifold reads:
\begin{equation}
S_G \,=\, -\kappa_2 N_2 + \kappa_4 N_4 \;,
\label{pg}
\end{equation}
where $N_k$ denotes the number of $k$-simplexes. The second part is
(for one copy of the gauge field):
\begin{equation}
S_M \,=\, \sum_{t_{abc}} o(t_{abc}) \left[ A(l_{ab}) + 
A(l_{bc}) + A(l_{ca})\right]^2 \;,
\label{sm}
\end{equation}
where $A(l_{ab})$ is a non-compact $U(1)$ gauge field living on 
link $l_{ab}$ and $A(l_{ab})$~=~$-A(l_{ba})$. The sum extends 
over all triangles $t_{abc}$ of the random lattice and 
$o(t_{abc})$ denotes the order of the triangle
$t_{abc}$, i.e.\ the number of simplexes sharing this triangle. 
We introduce $f$ copies of gauge fields.

\begin{figure} 
\epsfxsize=4in \centerline{ \epsfbox{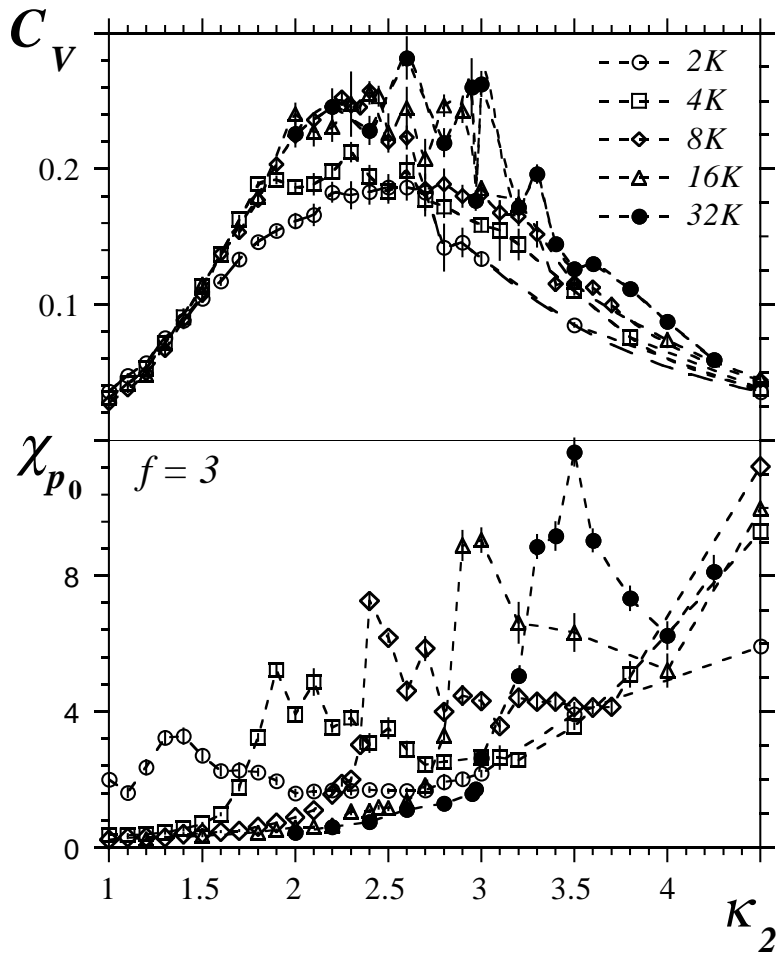}}
\caption[fig1]{\small The specific heat, $C_V$, and the
 susceptibility of the order of the most singular vertex,
 $\chi_{p_0}$, for three copies of gauge fields
 coupled to simplicial gravity.  This is for $N_4$ up to 32K.}
\label{fig1}
\end{figure}

We work in a pseudo-canonical ensemble of (spherical) manifolds, 
with almost fixed $N_4$. The model is defined by the partition function:
\begin{equation}
Z(\kappa_2, \bar{N}_4) \,=\,  \sum_{T} W(T) \; \int' 
\prod_{l \in T} dA(l) \; \; {\rm e}^{\textstyle - S_G - S_M - 
\delta (N_4- \bar{N}_4)^2 } .
\label{part}
\end{equation}
The sum is over all distinct triangulations $T$ and $W(T)$ is 
the symmetry factor taking care of equivalent re-labelings of 
vertexes. The prime indicates that the zero modes of the gauge 
field are not integrated. As is well known, one must allow the 
volume to fluctuate. The quadratic potential term added to the 
action ensures, for an appropriate choice of $\delta$, that these 
fluctuations are small. For more details see Ref.~\cite{us2}.

The result of our Monte Carlo simulations of the model defined by
Eq.~(\ref{part}), for 3 copies of gauge fields, is summarized in 
Figure~1.  There we plot both the specific heat: 
$C_V = \kappa_2^2 (\langle N_0^2\rangle 
  - \langle N_0\rangle^2)/ N_4$,
and the susceptibility of the order of the most singular vertex $p_0$:
$\chi_{p_0} = (\langle p_0^2\rangle - \langle p_0\rangle^2)/ N_4$.
This is done for $N_4$ up to 32K.  
The specific heat has a peak at $\kappa_2 \approx 2.5$;
the location of the peak is stable but its height
appears to saturate as the volume is increased.
This could indicate either a continuous phase transition, of
3rd or higher order, or simply a cross-over behavior.
The susceptibility $\chi_{p_0}$ also has a peak, rising with
the volume, signaling a change in the geometry.  Its location, however,
moves towards larger and larger values of the
coupling constant $\kappa_2$.
As collecting data at 32K already required a considerable effort,
going to a significantly larger volume is, for the moment,
out of question. Thus we cannot tell whether the
the location of the peak will tend to a finite value of $\kappa_2$
or will go to infinity.

The apparent inconsistency in the behavior of those two
geometric observables --- their fluctuations seem to be uncorrelated ---
is rather worrying.
A pessimistic view is that this implies the absence of a
true phase transition in the model; there will be no singularity
in the infinite volume limit, and the differences in scaling behavior
of the baby universe distribution
and $p_0$ in comparison to the crumpled phase, which we have
observed and which indicate a new phase, disappear in
the thermodynamic limit. While our simulations cannot
rule this out, another plausible explanation emerges as
the phase diagram is explored in more details as we do
in the next section.  We will return to this discussion
in the last section.

\section{Modified measure}

\noindent
The physical variables in the partition function Eq.~(\ref{part}) 
are not the gauge fields, but the plaquette values.  Let us replace
in our model, for the moment without any justification, the integration
over fields by the integration over plaquettes, 
which are by the same token
assumed to fluctuate independently.  They can then be 
integrated out, giving rise to a measure factor
\begin{equation}
M(T) \;\sim\; \prod_{j=1}^{N_2} o(t_j)^\beta
\label{me}
\end{equation}
with $\beta = -f/2$. This suggests to compare the two models, the one 
defined in the preceding section and that with the measure factor given 
by Eq.~(\ref{me}). We have done that leaving the parameter $\beta$ free, 
since it is likely that the estimate $\beta = -f/2$ is 
correct, if at all, in the limit $f \to \infty$ only.

\begin{table}
{\small
\caption{\small The number of different graphs $N_g$, for 
 volumes $N_4 =$ 32, 34, 36 and 38, and the corresponding 
 weights $W_f(N_4,N_0)$.
 This is shown both for pure gravity, and one and three gauge fields
 coupled to gravity ($f = 0$, 1, and 3).  All weights are
 normalized with the value at $N_4 = 6$.}
\vspace{8pt}
\begin{center}
\begin{tabular}{|rr|rrll|} \hline
$N_4$ & $N_0$  & $N_g$ & $W_0$  & $W_1$ & $W_3$\\  \hline

\vspace{-8pt}  & & & & &      \\
32 & 10 &   3886 &   2351430 & $0.002787853\ldots$
                          & $4.05397\ldots\times10^{-21}$ \\
   & 11 &   5943 &   3327045 & $0.003406732\ldots$
                          & $3.76146\ldots\times10^{-21}$ \\
   & 12 &   1700 & 6538455/8 & $0.000717216\ldots$
                          & $5.88670\ldots\times10^{-22}$ \\

\vspace{-8pt}  & & & & &      \\ 
34 & 10 &  11442 &   7502430 & $0.002104616\ldots$
                          & $1.72930\ldots\times10^{-22}$ \\
   & 11 &  26337 &  16396680 & $0.003873112\ldots$
                          & $2.27034\ldots\times10^{-22}$ \\
   & 12 &  13231 &   7545780 & $0.001504522\ldots$
                          & $6.40475\ldots\times10^{-23}$ \\
   & 13 &    922 &    411255 & $0.000069655\ldots$
                          & $2.14966\ldots\times10^{-24}$ \\ 

\vspace{-8pt}  & & & & &      \\
36 & 10 &  27765 &  18929925 & $0.001231989\ldots$
                          & $5.40586\ldots\times10^{-24}$ \\
   & 11 & 112097 &  74395157 & $0.004129758\ldots$
                          & $1.34373\ldots\times10^{-23}$ \\
   & 12 &  85734 &  54240610 & $0.002520113\ldots$
                          & $5.88459\ldots\times10^{-24}$ \\
   & 13 &  15298 & 26228930/3 &$0.000339780\ldots$
                          & $5.61994\ldots\times10^{-25}$ \\

\vspace{-8pt}  & & & & &      \\  
38 & 10 &  71295 &  50097510 & $0.000793825\ldots$
                          & $2.07384\ldots\times10^{-25}$ \\
   & 11 & 458083 & 315706725 & $0.004169780\ldots$
                          & $7.73454\ldots\times10^{-25}$ \\
   & 12 & 490598 & 328515075 & $0.003578606\ldots$
                          & $4.59400\ldots\times10^{-25}$ \\
   & 13 & 153773 &  97507410 & $0.000873276\ldots$
                          & $7.72935\ldots\times10^{-26}$ \\
   & 14 &   6848 &   3781635 & $0.000028143\ldots$
                          & $1.72338\ldots\times10^{-27}$ \\  \hline
\end{tabular}
\end{center}
}
\end{table}

First we use the strong coupling series. We have extended the
calculation presented in Ref.~\cite{us2}\footnote{There are typos
in Table~1 of Ref.~\cite{us2}:  The number of triangulations
of volume 24 with 9 vertexes is 34, not 13. Also, 
the weight $W_0$ of triangulations
of volume 28 with 11 vertexes is 77057$\;\frac{1}{7}$~=~539400/7,
not 77057.} up to $N_4 =38$. The new
results are summarized in Table~1. In Figure~2 we show a plot of 
the susceptibility exponent $\gamma$ versus $\beta$, calculated in the
large--$\kappa_2$ phase ($\kappa_2 = 10$) using the ratio method 
\cite{ratio}.  The values of $\gamma$ are extracted assuming 
a large-volume behavior of the canonical partition  
function: $Z(N_4) \sim \exp (\mu_c N_4) N_4^{\gamma -3 }$,
where $\mu_c$ is the critical cosmological constant.
We observe a range of (negative) couplings $\beta$ where 
a reliable estimate of $\gamma$ is obtained, converging
as more terms of the series are included
in the analysis. And, just as for the gauge field model,  
in this interval of $\beta$ the exponent  $\gamma$ decreases 
continuously from the branched polymer value, 
$\gamma = 1/2$, to a large negative value.

\begin{figure}
\epsfxsize=4in \centerline{ \epsfbox{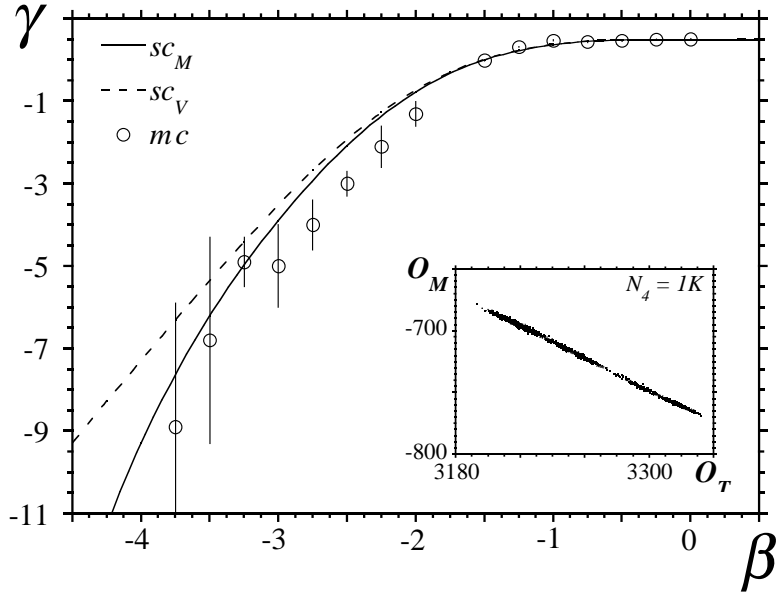}}
\caption[fig2]{\small Variations of $\gamma$ with both 
 the coupling $\beta$ to the measure term Eq.~(\ref{me})
 (solid line), and with the number of gauge fields $f$ (dashed line),
 calculated from the series expansion.  
 The value of $f$ has been rescaled using the relation Eq.~(\ref{rel}).
 Also shown are values of $\gamma$ measured in Monte Carlo
 simulations with the modified measure 
 for $\kappa_2 = 3$, varying $\beta$, on volume $N_4 = 4K$. 
 ({\it Insert}: The correlations between the effective
 actions stemming from integrating out the gauge fields
 and from the measure term, $O_M$ and $O_T$.)} 
\label{fig2}
\end{figure}

The similarity of results obtained with the two models,
gauge fields and modified measure, is in fact more spectacular.
In Figure~2 we also trace $\gamma$ versus $f$, making the
identification
\begin{equation}
\beta \;=\; - \frac{f}{2} -  \frac{1}{4}.
\label{rel}
\end{equation}
In an interval of the values of $\beta$ the two curves coincide!
Actually, there is close agreement in the parameter range
where the ratio method seems to give reliable values
of $\gamma$. Note, that the leading term in Eq.~(\ref{rel})
coincides with our earlier estimate. Of course, Eq.~(\ref{rel})
is meaningless for too small values of $f$ ($-\beta$):
there $\gamma = 1/2$ for both models and the mapping is trivial.

We have further checked this remarkable equivalence in a 
numerical experiment, measuring the effective change 
to the action Eq.~(\ref{pg}) stemming from these two
seemingly unrelated modifications of the standard model.  
In a Monte Carlo simulation of pure gravity, with 
$\kappa_2 = \beta = f = 0$ and $N_4 = 1K$, 
we measured the effective actions: 
\begin{equation}
O_M \;=\; \sum_j \log (o(t_j))  \qquad {\rm and} \qquad
O_V \;=\; \frac{1}{2} \;
    \log \left ( \frac{\pi^{N_1-N_0+1}}{\Delta} \right ),
\end{equation} 
where $\Delta$ is the determinant
associated with the integration over one species of gauge fields.
We observe a very strong linear correlation between those two
observables, as shown in the insert in Figure~2.  
A linear fit yields a slope $-0.65$, roughly 
compatible with Eq.~(\ref{rel}). 

Further corroboration comes from Monte Carlo simulations at
$\beta=-1.75$ (corresponding to $f=3$ in Eq.~(\ref{rel})), 
for $N_4=$~4K and 8K and for $\kappa_2$ ranging from 0 to 4.5 .
We do not wish to drown the reader in figures. It suffices 
to say that, up to a shift in $\kappa_2$ not larger than $\sim 0.5$, 
the results concerning the first two cumulants of $N_0$ and 
the orders of the three most singular vertexes are practically 
the same\footnote{The shift in $\kappa_2$ depends, however, on
the observable.}.

\begin{figure}
\epsfxsize=4in \centerline{ \epsfbox{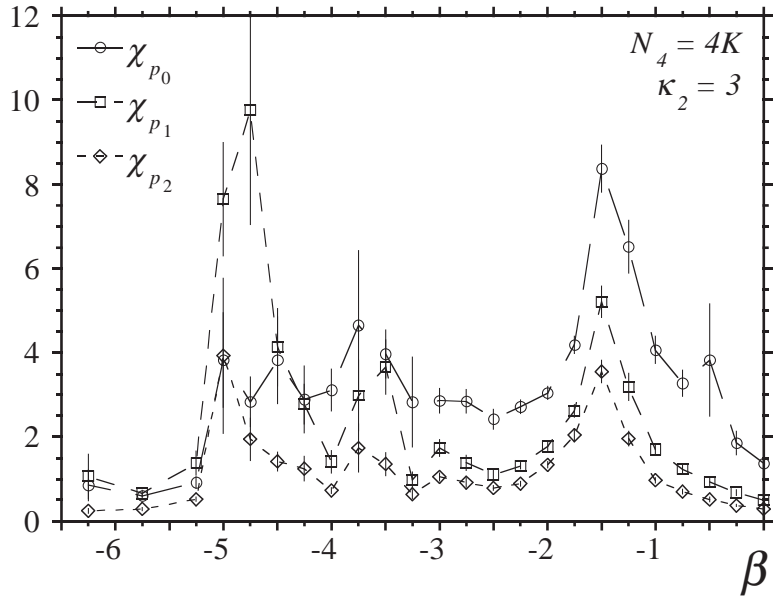}}
\caption[fig3]{\small The fluctuations in the order of the 
 three most singular vertexes, $p_0$, $p_1$ and $p_2$,
 for simplicial gravity with a modified measure.  This
 is for $\kappa_2 = 3$ and volume is $N_4 = 4K$.}
\end{figure}

As simulating the model with a modified measure is less  
CPU-demanding than working with several copies of gauge fields,
it has been possible to explore the phase diagram in more
details.  In addition, the coupling $\beta$
can be varied continuously, whereas we
are restricted to integer values of $f$.
We did simulations for 
$\kappa_2 = 3$, $N_4 = 4K$, and for several values of $\beta$.  
We observe strong fluctuations in geometry {\it both} at 
$\beta \approx -1.5$ and $\beta \approx -5$.
This is indicated by peaks in the susceptibility of the 
various geometric observables; in Figure~4 we show
this for the orders of the three most singular vertexes.
The first signal corresponds roughly to the transition from
branched polymers to the crinkled phase, as indicated by the series 
expansion, whereas the latter is a transition to a 
crumpled phase for $\beta \gtrsim -5$.  At the moment we have
only investigated this at one lattice volume, obviously 
further exploration is needed to establish that 
this indeed corresponds to two {\it distinct} transitions.

To end this section let us mention that this universality in the
back-reaction on the geometry also holds for other 
modifications of the standard model.  Using the series expansion
we have investigated the effects of adding:
({\it a}) Gaussian scalar fields,
({\it b}) a discretized $R^2$-term as used in Ref.~\cite{jan},
and ({\it c}) a modified measure using a product of vertex orders.
In all three cases it is possible, by an appropriate rescaling
of the corresponding couplings, to map the extracted values
of $\gamma$ on the curve in Figure~2, although the agreement is
not as spectacular as that between the model with gauge fields
and the one with a measure modified according to 
Eq.~(\ref{me}), respectively.

\section{The nature of crinkled manifolds}

We have explored further the nature of the hypothetical
crinkled phase in Monte Carlo simulations.  We have calculated
the exponent $\gamma$, from the distribution of baby
universes, along two lines in the phase 
diagram: $l_1 = \{\kappa_2 = 3,\beta\}$ 
and $l_2 = \{\kappa_2,\beta=-3.5\}$. 
The values of $\gamma$ measured along $l_1$,
included in Figure~2, agree reasonably with 
the predictions of the series expansion.  
The values of $\gamma$ measured along $l_2$ are shown in Table~2.  
For small values of $\kappa_2$ it is
not possible to extract any reliable value. This is to be
expected as the model is in the crumpled phase.  
For $\kappa_2 \gtrsim 3.5$, however,
we get a consistent value $\gamma \approx -4.5$,  
compared with the prediction of the strong 
coupling expansion, $\gamma \approx -6.2$, obtained
for $\beta = -3.5$ and $\kappa_2$ large.  

\begin{table}
\begin{center}
\caption[tab3]{\small Values of the string
 susceptibility exponent $\gamma$, measured in
 Monte Carlo simulations at $N_4 = 4K$, for $\beta = -3.5$
 and varying $\kappa_2$.  The series expansion predicts
 $\gamma \approx -6.2$, for $\beta = -3.5$ and large
 values of $\kappa_2$.}
\vspace{8pt}
\begin{tabular}{|c|c||c|c|} \hline
$\kappa_2$  &  $\gamma$  &  $\kappa_2 $  &  $\gamma$  \\
\hline
3.5     &  -5.75(15)   &   6.0    &   -4.71(46)   \\
4.0     &  -6.24(26)   &   6.5    &   -4.19(30)   \\
4.5     &  -6.23(23)   &   7.0    &   -4.27(38)   \\
5.0     &  -4.41(12)   &   7.5    &   -4.14(70)   \\
5.5     &  -4.42(25)   &          &               \\  \hline
\end{tabular}
\end{center}
\end{table}

Other exponents characterizing the fractal geometry are
the Hausdorff and spectral dimensions of the manifolds, 
$d_H$ and $d_s$.
The former is related to the volume of space
within a sphere of geodesic radius $r$ from a marked point: 
$v(r) \sim r^{d_H}$, whereas the spectral dimension defines
the return probability for a random walker on the
triangulation: $p(t) \sim t^{-d_s/2}$ \cite{spec1}. 
The time $t$ is measured in units of jumps between 
neighboring vertexes, with hopping probability given
by the inverse of the coordination number\footnote{
In calculating both the Hausdorff and spectral dimensions
we used the dual graph. This is more natural as we measure 
at fixed volume $N_4$, i.e.\ fixed number of vertexes in the dual
graph.}. 
On smooth regular manifolds those two definitions of dimensionality
coincide.  However, on highly fractal manifolds, like the
ones that dominate the partition function Eq.~(\ref{part}),
they are in general different.

We have extracted the Hausdorff dimension for $\kappa_2 = 4.5$
and $\beta = -1.75$ and $-3.5$, from the expected scaling
behavior of the average distance between two simplexes:  
$\langle r_{ij} \rangle_{N_4} \sim N_4^{1/d_H}$.  
Measurements at volume $N_4$~=~4K to 32K were included in the fit. 
For $\beta = -1.75$ we got $d_H = 3.57(16)$, which should 
be compared to $d_H = 3.97(15)$ quoted in Ref.~\cite{us2}
for $f=3$ and $\kappa_2 = 4.5$.  For $\beta = -3.5$ we got
$d_H \approx 5$, but this estimate is less reliable.

\begin{table}
\begin{center}
\caption[tab3]{\small The extracted values of the spectral dimension
 $d_s$ for simplical gravity with a modified measure.  This is 
 for $\kappa_2 = 4.5$ and both for $\beta = 0$
 (branched polymer phase) and for $\beta = -1.75$ and -3.5
 (crinkled phase).}    
\vspace{8pt}
\begin{tabular}{|c|c|c|c|} \hline
$N_4$  &  $\beta = 0.00$  & $\beta = -1.75$  &  $\beta = -3.5$
 \\ \hline
2K     &   1.33(1)  & 1.50(1)    &   1.77(3)   \\
4K     &   1.33(1)  & 1.51(1)    &   1.80(5)    \\
8K     &   1.33(1)  & 1.51(2)    &   1.77(4)   \\
16K    &   1.33(2)  & 1.52(3)    &   1.77(4)   \\  \hline
\end{tabular}
\end{center}
\end{table}

The measured values of the spectral
dimension at $\kappa_2=4.5$ for $N_4$ ranging from 2K to 16K
and for three values of $\beta$ are shown in Table~3.  
In the branched polymer phase, at $\beta = 0$, we get
$d_s \approx 1.33$, to be compared to the theoretical value for
generic branched polymers: $d_s = 4/3$ \cite{spec1}.
In the crinkled phase, on the other hand, we get values for the
spectral dimension significantly larger than
$4/3$ and which, moreover, seem to increase as $\beta$ is 
decreased: $d_s \approx 1.5$ at $\beta = -1.75$, and
$d_s \approx 1.75$ at $\beta = -3.5$.  In all cases, the
values obtained at different volumes agree within the
numerical accuracy. 

\section{Discussion}

\noindent
The quasi-equivalence between the model with gauge fields and that with 
modified measure cannot be a coincidence. The simplest 
explanation is that the correlation between plaquettes falls
rapidly with the distance on the lattice. This sheds a new light 
on the results of Ref.~\cite{us2}. The effective action used in
the speculation of Ref.~\cite{jk} is derived from trace anomalies, 
assuming conformal invariance. This regime is apparently different 
from the one we observe on our disordered lattice.
If so, then the mechanism of suppression of polymerization 
observed in the model is not the one which has been 
expected and a faithful implementation of the idea of
Ref.~\cite{jk} remains an open problem.  

We are now in a better position to discuss 
certain points which were left obscure in Ref.~\cite{us2}. 
In particular, mean-field arguments give perhaps some insight into
our results:

Following Ref.~\cite{bb} define a mean-field 
model by the canonical ensemble 
partition function
\begin{equation}
z \;\sim\; \sum_{N_0 \leq N_4/4} \; e^{\; \textstyle \kappa N_0} 
\;\sum_{P} \; \prod_{q \in P} q^\alpha \; 
\delta \left (\sum_{q \in P} q - 5 N_4 \right )
\; , \qquad \alpha < -2,
\label{zmf}
\end{equation}
where $P$ denotes partitions of the integer $5 N_4$ 
into $N_0$ ``vertex orders'' $q$  and $\alpha$ is an adjustable 
parameter\footnote{It might appear more natural here to assume that
triangle and not vertex orders are the variables
fluctuating independently (up to the global kinematic
constraint). It turns out that one would then obtain an
unphysical model, predicting singular triangles in the crumpled 
phase, in variance with observations.}. At large $N_4$ 
one can write
\begin{equation}
z \;\sim\; \int_0^{\frac{1}{4}}  {\rm d}r \;
e^{ \textstyle \;  N_4 [\kappa r + f(r)]} \;, \qquad r = N_0/N_4 \; ,
\label{mf}
\end{equation}
the function $f(r)$ having two remarkable properties:
$f'(r) = $const (call it $- \kappa_{cr}$) for $r < r^*$, and 
$f''(r) < 0$ for $r > r^*$. Furthermore $r^*$ increases with
$-\alpha$ (see Ref.~\cite{bb} for details). 

Thus, in the thermodynamic limit and as long as 
$r^* < 1/4$, the model has the following 
two phases: for $\kappa < \kappa_{cr}$
one has $r=0$, while for $\kappa > \kappa_{cr}$ one finds $r=r_{sp}$,
where $r_{sp} \geq r^*$ is the position of the saddle-point. 
The ``latent heat'' at the transition equals $r^*$. Loosely speaking,
the two phases correspond to crumpled and branched 
polymer geometries, respectively: For $\kappa < \kappa_{cr}$ a few
vertexes have orders $\sim N_4$, while for $\kappa > \kappa_{cr}$
vertex orders are bounded.

The situation changes when $r^* > 1/4$: there is 
no saddle point in the integration range 
and at $\kappa = \kappa_{cr}$ the system jumps from $r=0$ to 
$r = 1/4$ (this argument, implicit in 
Ref.~\cite{bb}, is emphasized in 
Refs.~\cite{syr,bb2}). In the model, there are still singular
vertexes (i.e. with order $\sim N_4$) 
when $\kappa > \kappa_{cr}$, but the order of the
most singular vertex drops suddenly as one moves across
$\kappa = \kappa_{cr}$. Hence, at  $\kappa = \kappa_{cr}$
the corresponding susceptibility is expected to have a peak
with height $\propto N_4$.

The overall picture shows some similarity to what
we observe, especially if one recalls
the results mentioned at the end of 
Section~3, implying that the increase of $-\alpha$ can mimic the 
increase of $f$ or $-\beta$ in the modified simplicial
gravity model.

Although the mean field model clearly helps understanding
our results, it should perhaps not be taken too literally.
In particular, the model predicts that the
only coupling space region where the crinkled phase survives in the 
thermodynamic limit is the region where $N_0/N_4 = 1/4$, i.e.
where the naive Regge curvature sticks to its upper kinematic
bound. This scenario is plausible. If it is true, 
the crinkled phase is an unlikely candidate for
a physical phase of quantum gravity. 
In the data (see Figure~1) the transition 
point seems to run towards large values of $\kappa_2$, where 
indeed $\langle N_0 \rangle/N_4 \to 1/4$. However, one 
observes $N_0/N_4 = 1/4$ also in the branched polymer phase, 
at finite volume and large $\kappa_2$. This could be just a finite size
effect. We have no evidence for a dramatic jump
of $\langle N_0 \rangle/N_4 $ as the system moves from the crumpled 
to the crinkled phase. Actually, the behavior of the specific heat shown
in the upper part of Figure~1, seems to exclude any 1st order transition.
Also, the most singular vertex susceptibility increases slower than
the volume.

We should also mention another point where the mean field model
appears in variance with data. As already mentioned, the
qualitative prediction of the model is that the latent heat
{\em increases} with $-\alpha$ (or $-\beta$, or $f$).
However, in Ref.~\cite{us2} we have noted that the latent heat at 
the transition point {\em decreases} (by a factor of 
2 at 32K) as one moves $f$ from 0 to 1. This observation 
is compatible with the claim made in Ref.~\cite{hotta}, but in 3d, 
that the transition becomes of second order when the power in 
the measure (in their case the power of the vertex order) 
becomes sufficiently negative. 

\begin{figure}
\epsfxsize=4in \centerline{ \epsfbox{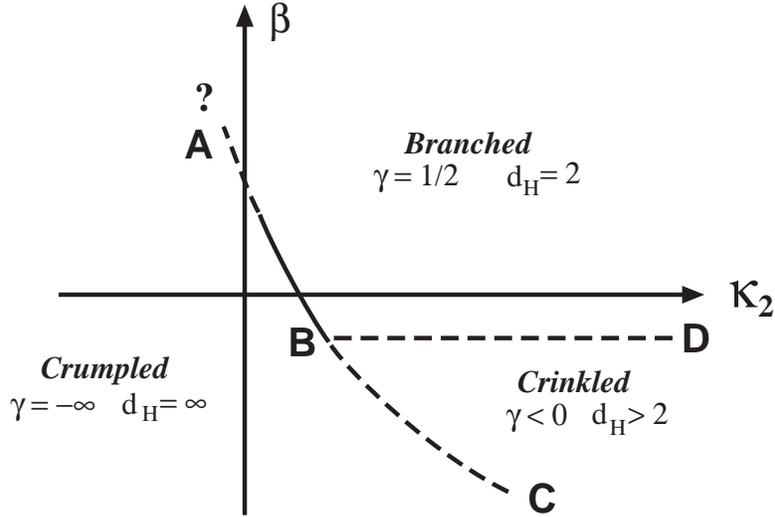}}
\caption[fig5]{\small A schematic phase diagram of simplical
 gravity modified with the measure Eq.~(\ref{me}).  At least along
 a portion of the line {\sc \bf BA} the phase 
transition is discontinuous.
 The same phase diagram should hold for
 gauge fields coupled to gravity, Eq.~(\ref{part}), replacing
 $\beta$ by the number of fields $f$ 
 rescaled by the relation (\ref{rel}).}   
\label{fig5}
\end{figure}

We believe that the results of our study 
can be tentatively summarized by the 
phase diagram\footnote{This phase diagram suggests an 
explanation for the inconsistency
observed in Section~2 in the simulations with 3 gauge fields.
As discussed in Section~3, the transition from the branched
polymer phase to the crinkled phase occurs at
$\beta \approx -1.5$. This is accompanied by strong fluctuations
in geometric observables such as the maximal order
of vertexes.  By the relation Eq.~(\ref{rel}) this value of $\beta$
corresponds to $f \approx 2.5$. Hence simulating 3 gauge fields
might have been an unlucky choice. It it possible that
varying $\kappa_2$ we have followed the line
{\sc \bf BD} in the phase diagram,
uncomfortably close to the transition region.}
of Figure~5.  There are three distinct regions corresponding to 
the branched polymer phase, the 
crumpled phase and the hypothetical crinkled  
phase, respectively.  The solid line {\sc \bf BA} represents the line
of phase transitions separating the crumpled and
the branched polymer phases.  The two dashed lines, {\sc \bf BC}
and {\sc \bf BD}, separate the crinkled phase from, respectively,
the crumpled and branched polymer ones.  It is unclear yet
whether these lines represent genuine phase  
transitions, in which case the point {\sc \bf B} is a tricritical
point, or a cross-over behavior.
Further simulations are needed in order to clarify this
issue and to verify that this phase structure
survives in the thermodynamic limit.
 
\vspace{20pt}
\noindent
{\bf Acknowledgments:}
We are indebted to P. Bialas for discussions and to
B. Klosowicz for help. We have used the computer
facilities of the CRI at Orsay
and of the CNRS computer center IDRIS, the HRZ, Univ.\
Bielefeld and HRZ Juelich.
A part of the simulations were carried out
at the IBM SP2 at SARA.
S.B.\ was supported by FOM.
Z.B.\ has benefited from
KBN grants 2P03B19609 and 2P03B04412. 
B.P.\ is grateful to the Center for Computational Physics
at the University of Tsukuba for kind hospitality during
the final part of this investigation.
J.T.\ was supported
by the DFG, under the contract PE340/3-3, and
G.T.\ by the Humboldt Foundation.

\end{document}